\documentclass[11pt]{article} \textwidth 170mm \textheight 225mm
\usepackage{amsfonts}
\usepackage{amssymb}
\usepackage{dcolumn}
\usepackage{amsthm}
\usepackage{amsmath}
\usepackage{multicol}
\usepackage{newlfont}
\usepackage{newlfont}
\usepackage{multicol}
\usepackage{bm}
\usepackage{txfonts}
\usepackage{graphics}
\usepackage{graphicx}
\usepackage{color}
\usepackage{indentfirst}
\begin{document}

\title{First-principles calculations on finite temperature elastic properties of B2-AlRE (RE=Y, Tb, Pr, Nd, Dy) intermetallics  \footnote{The work is supported by the National Natural Science Foundation
of China (11074313). }}
\author{Rui Wang\footnote{Tel: +8613527528737; E-mail: rcwang@cqu.edu.cn.}, Shaofeng Wang, and Xiaozhi Wu\\
{\small  { Institute for Structure and Function and
Department of physics, Chongqing University, }}\\ {\small {Chongqing 400044, People's Republic of China. }} }

\date{}

\maketitle
\begin{abstract}
\baselineskip 18pt
\noindent  We have investigated the finite temperature elastic properties of AlRE (RE=Y, Tb, Pr, Nd, Dy) with B2-type structures from first principles. The phonon free energy and thermal expansion is obtained from the quasiharmonic approach based on density-functional perturbation theory. The static volume-dependent elastic constants are obtained from energy-strain functions by using the first-principles total-energy method. The comparison between our predicted results and the ultrasonic experimental data for a benchmark material Al provides excellent agreements. At $T=0K$, our calculated values of lattice equilibrium volume and elastic moduli of our calculated AlRE (RE=Y, Tb, Pr, Nd, Dy) intermetallics agree well with the previous theoretical results. The temperature dependent elastic constants exhibit a normal behavior with temperature, i.e., decrease and approach linearity at higher temperature and zero slope around zero temperature. Furthermore, the anisotropy ratio and sound velocities as a function of temperature has also been discussed.
\end{abstract}

\vskip 0.1in{\small   } \vskip 0.2in

\vskip 0.1in

 \vskip 0.3in

\baselineskip 20pt


\section{Introduction}

Aluminium alloys are widely used in the automotive, aircraft and aerospace industries, and interest of their applications is rapidly increasing.  Addition of rare earth (RE) elements to Al-based metal alloys have received great attention since it can result in increasing their wear resistance \cite{Shi2010}, mechanical properties \cite{Knipling},
electrochemical behavior \cite{Rosalbinoa} and thermal stability \cite{Nie2002},etc.  The development and design of new Aluminum alloys need more fundamental physical data of these alloys. Recently, some investigations focusing on AlRE, Al$_{2}$RE, and Al$_{3}$RE systems intensely are performed from first-principle calculations based on density functional theory (DFT) \cite{Srivastava,Tao1,Tao,Asta,Ugur,Pang2011,Zhou2010,Wang2011s}. The mechanical and thermodynamical properties of B2-AlRE intermetallics have been investigated from using the projector augmented wave (PAW) method within generalized gradient approximation (GGA) by Tao and Ouyang et al \cite{Tao,Tao1}, while Srivastava et al \cite{Srivastava} also investigated the electronic and thermal properties using the tight binding linear muffin tin orbital (TB-LMTO). These results are very important to the novel material design and further scientific and technical investigations.

Elastic properties of a solid are very important because some physical properties such as the bulk modulus, shear modulus,
Young¡¯s modulus, anisotropy, and the velocity of sound  can be  determined from its elastic constants \cite{Ravindran}. Furthermore, the finite temperature elastic constants are critical for predicting and understanding the mechanical strength, stability, and phase transitions of a material \cite{Gulsern}. Though modern electronic-structure methods based on DFT can compute the zero-temperature pressure dependence elastic moduli from first principles accurately \cite{Karki}, treating the corresponding temperature dependence of elastic constants is still a formidable challenge \cite{Orlikowski}. However, several groups have also attempted to investigate the temperature-dependent elastic properties  by using the first-principles DFT methods \cite{Orlikowski,Kadas,Sha,Wang,Shang}. More recently, we have investigated the temperature-dependent elastic properties of two ductile rare-earth intermetallics YAg and YCu, and MgRE intermetallics from using the first principles approach successfully \cite{Wang2011-1, Wang2011-2}. So far as we know, the temperature-dependent elastic moduli of the novel B2-tpye structures AlRE intermetallics have not been reported in the literature. We investigate the finite temperature elastic properties for AlRE (RE=Y, Tb, Pr, Nd, Dy) using a first-principles method \cite{Wang,Shang} in present study. The elastic constants as a function of temperature, the anisotropy ratio, and the temperature-dependent sound velocity have been discussed.

\section{Theoretical methods}

The isothermal elastic constants are derived from the second-order strain derivative of the Helmhotz free energy  \cite{Brugger}
\begin{equation}\label{Cijkl}
C_{ijkl}^{T}=\frac{1}{\Omega}\frac{{\partial}^{2}F}{\partial \varepsilon_{ij}
\partial \varepsilon_{kl}}\bigg|_{T,\varepsilon=0},
\end{equation}
where $\Omega$ is equilibrium volume at temperature $T$, $\varepsilon$ denotes the linear strain tensor, and the Helmholtz free energy $F$ can be expressed as the function of volume $\Omega$ and temperature $T$,
\begin{equation}\label{Ftotal}
F(\Omega, T)=E_{\mathrm{0}}(\Omega)+F_{\mathrm{el}}(\Omega,T)+F_{\mathrm{ph}}(\Omega, T),
\end{equation}
where $E_{\mathrm{0}}$ is the zero-temperature total energy of the electronic ground state, $F_{\mathrm{el}}$ represents the electron-thermal contribution from finite temperature electrons, and $F_{\mathrm{ph}}$ denotes the phonon free energy arising from the lattice vibrations. For the electron-thermal free energy, an acceptable approximation to calculate it is to use the Mermin statistics \cite{Wang2000-1, Wang2000-2} through $F_{\mathrm{el}}=E_{\mathrm{el}}-TS_{\mathrm{el}}$. Within the mean-filed approximation, the thermal electronic energy $E_{\mathrm{el}}$  and  entropy $S_{\mathrm{el}}$ can be give by $E_{\mathrm{el}}=\int \epsilon n(\epsilon)f(\epsilon) d\epsilon-\int^{\epsilon_{F}}\epsilon n(\epsilon) d\epsilon$ and $S_{\mathrm{el}}=-k_{\mathrm{B}}\int n(\epsilon)[f(\epsilon)\ln f(\epsilon) +(1-f(\epsilon))\ln (1-f(\epsilon))] d\epsilon$, respectively, where $k_{B}$ is Boltzmann constant, $\epsilon_{F}$ is Fermi level, $f(\epsilon)$ is the Fermi-Dirac distribution, and $n(\epsilon)$ is the density of states. The phonon free energy can be obtained within the framework of quasiharmonic approach (QHA) by $F_{\mathrm{ph}}=  \sum_{\kappa}\Big[\frac{1}{2}{\hbar\omega_{\kappa}}+{k_{B}T}\ln\big(1-e^{-{\hbar\omega_{\kappa}}/{k_{B}T}}\big)\Big]$, where $\omega_{\kappa}$ represents an individual phonon frequency.

For the B2-type AlRE intermetallics with cubic crystal, there are only three independent constants $C_{11}^{T}$, $C_{12}^{T}$, and $C_{44}^{T}$ (in Voigt notation). A more convenient set for computations are $C_{44}^{T}$ and two linear combinations $B_{T}$ and $\mu_{T}$. The bulk modulus $B_{T}=(C_{11}^{T}+2C_{12}^{T})/3$ is the resistance to deformation by a uniform hydrostatic pressure and can be obtained from the the equation of state (EOS) \cite{Gulsern}, using the Vinet EOS \cite{Vinet} in the current calculation. The shear modulus $\mu_{T}=(C_{11}^{T}-C_{12}^{T})/2$ is the resistance to shear deformation across the (110) plane in the [1\={1}0].

We apply a  volume conserving tetragonal strain to calculate shear modulus $\mu_{T}$,
\begin{equation}\label{strain1}
\varepsilon(\delta)=(\delta, \delta, (1+\delta)^{-2}-1, 0, 0, 0),
\end{equation}
and the corresponding free energy  is
\begin{equation}\label{Fs1}
F(\Omega,\delta)=F(\Omega,0)+6\mu_{T} \Omega \delta^{2}+O(\delta^{3}),
\end{equation}
where $F(\Omega,0)$ is the free energy of the unstrained state.
$C_{44}^{T}$ is obtained from a volume-conserving monoclinic stain,
\begin{equation}\label{strain2}
\varepsilon(\delta)= (0, 0, \delta^2/(1-\delta^2), 0, 0, 2\delta)
\end{equation}
which leads to the free energy change
\begin{equation}\label{Fs2}
F(\Omega,\delta)=F(\Omega,0)+2C_{44}^{T} \Omega \delta^{2}+O(\delta^{4}).
\end{equation}
The elastic constants are calculated from the fitted quadratic coefficients of the strain free energy. Then, the elastic constants $C_{11}^{T}$ and $C_{12}^{T}$ are obtained from $\mu_{T}$ and $B_{T}$.

The electronic energy, including the static energy at ground state and thermal electronic contribution to the free energy, are computed by using the density functional theory (DFT).  For the first-principles DFT calculations, we employed the plane-wave basis projector augmented wave (PAW) method \cite{Blochl, Kresse4} in the framework of of the generalized gradient approximation (GGA) in the
Perdew-Burke-Ernzerhof form \cite{Perdew1,Perdew2} as implemented in the VASP
code \cite{Kresse1, Kresse2,Kresse3,Kresse5}.  Since high accuracy is needed to evaluate the elastic constants, the convergence of strain energies with respect to the Brillouin zone integration was carefully checked by repeating the calculations for $21\times21\times21$ and $25\times25\times25$  Monkhorst-Pack \cite{Monkhorst} meshes at equilibrium volume at T=0K, and we found that fluctuation both for $C_{44}^{T}$ and $\mu_{T}$ is lower than 0.5GPa. Hence, we used $21\times21\times21$ in the full Brillouin zone giving 726 irreducible k-points. In addition, we used a high plane-wave energy cutoff of 600eV which is sufficient to calculate the elastic moduli accurately.

To calculate the phonon free energy in Eq. (\ref{Ftotal}), we have carried out the density-functional perturbation theory (DFPT) within supercell approach. Real-space force constants of supercells are calculated in the DFPT implemented in the VASP code \cite{Kresse5}, and phonon frequencies are calculated from the force constants using the PHONOPY package \cite{Togo2010,Togo2008,Togop}.  The chosen supercell size strongly influences on the thermal properties. We give $F_{ph}$ at 300K and 1000K with $2\times2\times2$, $3\times3\times3$, and $4\times4\times4$ supercells for AlRE (RE=Y, Tb, Pr, Nd, Dy) in Table \ref{testsize}. We found that the phonon free energies between $3\times3\times3$ and $4\times4\times4$ are less that 0.5kJ/mol, so the adequate supercell size consisting of $3\times3\times3$ unit cells containing 54 atoms is used. The total Helmholtz free energy including the electronic energy and phonon free energy are calculated at 13 volume points. The equilibrium volume $\Omega_{0}(T)$ at temperature $T$ and the bulk modulus $B_{T}$ are obtained by minimizing Helmholtz free energy Eq. (\ref{Ftotal}) with respect to $\Omega$ from fitting the integral form of the Vinet equation of state (EOS) \cite{Vinet}.

The temperature-dependent elastic constants are obtained in the following three steps \cite{Wang,Shang}. The first step is calculating the  the equilibrium volume  $\Omega_{0}$ and thermal expansion as a function of temperature $T$  by using the first-principles quasiharmonic approximation (QHA), while isothermal bulk modulus $B_{T}$ is also obtained from fitting to the Vinet equation of state \cite{Vinet}. The detailed procedure of calculating the thermal expansion coefficient, isothermal bulk modulus $B_{T}$ were shown in our recent work \cite{Wang2011}. In the second step, we predict the static elastic constants as a function of volume $C_{ij}^{T}(\Omega_{0})$ by using the energy-strain relation based on Eqs.(\ref{strain1}-\ref{Fs2}). In the third step, the calculated elastic constants from the second step at the volume $\Omega_{0}(T)$ are approximated as those at finite temperatures, i.e., $C_{ij}^{T}(T)=C_{ij}^{T}[T(\Omega_{0})]$.

\section{Results and discussion}

In Table \ref{table} we list the equilibrium volume $\Omega_{0}$, elastic constants $C_{ij}$ at $T=0K$ for AlRE (RE=Y, Tb, Pr, Nd, Dy) in comparison with the previous calculated results obtained by Tao and Ouyang, et al.\cite{Tao1}. One can see that the values in our calculations are in excellent agreement with the previous DFT data \cite{Tao1}. It is noticeable that the requirement of mechanical stability for all calculated AlRE, namely,  $C_{11}-C_{12}>0$, $C_{11}>0$, and $C_{44}>0$ \cite{Nye}, is satisfied. In addition, the presented results of AlY and AlPr are also agree well with theoretical results obtained from the tight binding linear muffin tin orbital (TB-LMTO) by Srivastava et al \cite{Srivastava}.

\begin{figure}
\scalebox{1.0}[1.1]{\includegraphics{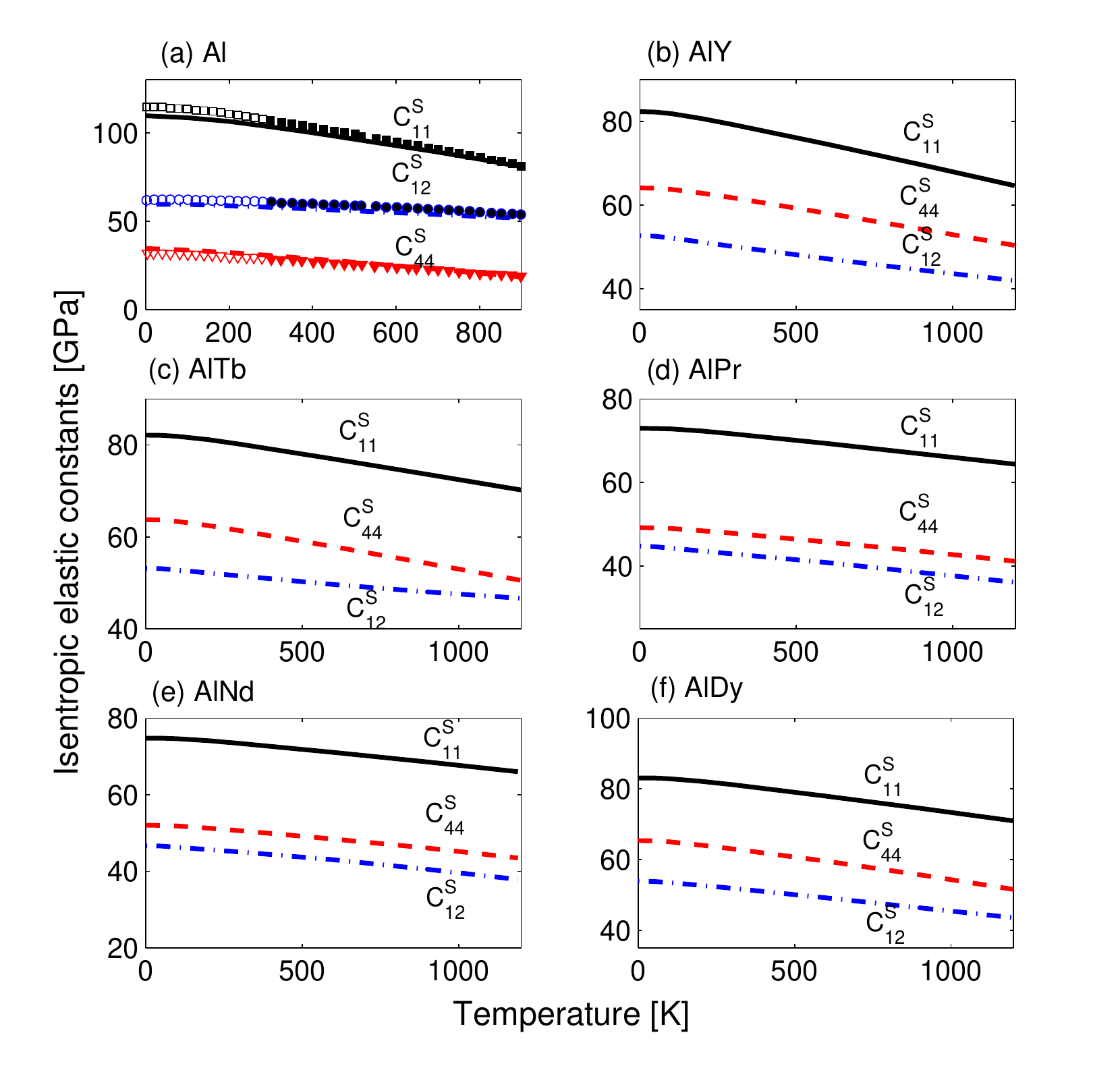}}
\caption{ The calculated isentropic elastic constants for (a) Al, (b) AlY, (c) AlTb, (d) AlPr, (e) AlNd, and (f) AlDy. The solid, dashed-dotted  and dashed curves represent the values of $C_{11}^{S}$, $C_{12}^{S}$ and $C_{44}^{S}$, respectively. For Al, the empty symbols denote the corresponding experimental data of Kamm et al \cite{Kamm}, and the full symbols indicate the experimental values from Gerlich et al \cite{Gerlich}.}
\label{elastic}
\end{figure}

The temperature-dependent elastic constants have not been reported theoretically and experimentally for all AlRE intermetallics, so we choose a benchmark material Al, which has the face-center-cubic (FCC) structure, accompanied by available experimental data to get a useful test of the accuracy of the method and the precision of our calculations. Furthermore, it is note that Eq. (\ref{Cijkl}) gives the isothermal elastic constants $C_{ij}^{T}$ \cite{Birch}, but most of the experimental data of elastic constants are usually measured as isentropic elastic constants. In order to compare with experimental data, the isothermal elastic moduli $C_{ij}^T$ must also be transformed to the isentropic elastic moduli by the following relation \cite{Davies}
\begin{equation}\label{TtoS}
C_{ij}^{S}=C_{ij}^{T}+\frac{T\Omega}{C_{\Omega}}\sum_{kk'}\alpha_{k}\alpha_{k'}C_{ik}^{T}C_{jk'}^{T}
\end{equation}
where $C_{\Omega}$ is the specific heat at constant volume and $\alpha_{k}$ is the linear thermal expansion tensor. For the cubic crystal, Eq. (\ref{TtoS}) simplifies to \cite{Orlikowski}
\begin{eqnarray} \label{CTS}
&C_{44}^{S}=C_{44}^{T},\nonumber \\
&C_{11}^{S}-C_{11}^{T}=C_{12}^{S}-C_{12}^{T}=\frac{T\Omega}{C_{\Omega}}\alpha^{2}B_{T}^{2},
\end{eqnarray}
with $\alpha$ is the volume thermal expansion coefficient, which is defined as $\alpha=(1/\Omega) \partial \Omega/\partial T$. For the temperature dependence of the isentropic elastic moduli for Al and AlRE (RE=Y, Tb, Pr, Nd, Dy) intermetallics, we present in Figure \ref{elastic} our calculated results. In Figure \ref{elastic}(a), we also show our current date of Al together with data taken from ultrasonic measurements \cite{Gerlich,Kamm}. The overall agreement of the present results with experiment is very good at all temperatures, even near melt. Our results also show no discrepancy with previous calculated values obtained  from using the first-principles approach by Wang et. al. \cite{Wang}. For all calculated AlRE intermetallics, we find the isentropic elastic constants $C_{11}^{S}$, $C_{12}^{S}$ and $C_{44}^{S}$  decrease with increasing temperature, since thermal expansion may soften the elastic moduli. We also find that the trend of $C_{11}^{S}$, $C_{12}^{S}$ and $C_{44}^{S}$  approach linearity at higher temperature and zero slop around zero temperature. The calculated curves from Figure \ref{elastic} indicate that the elastic constants of AlRE follow a normal behavior with temperature\cite{Wang,Orlikowski}.

\begin{figure}
\scalebox{0.7}[0.7]{\includegraphics{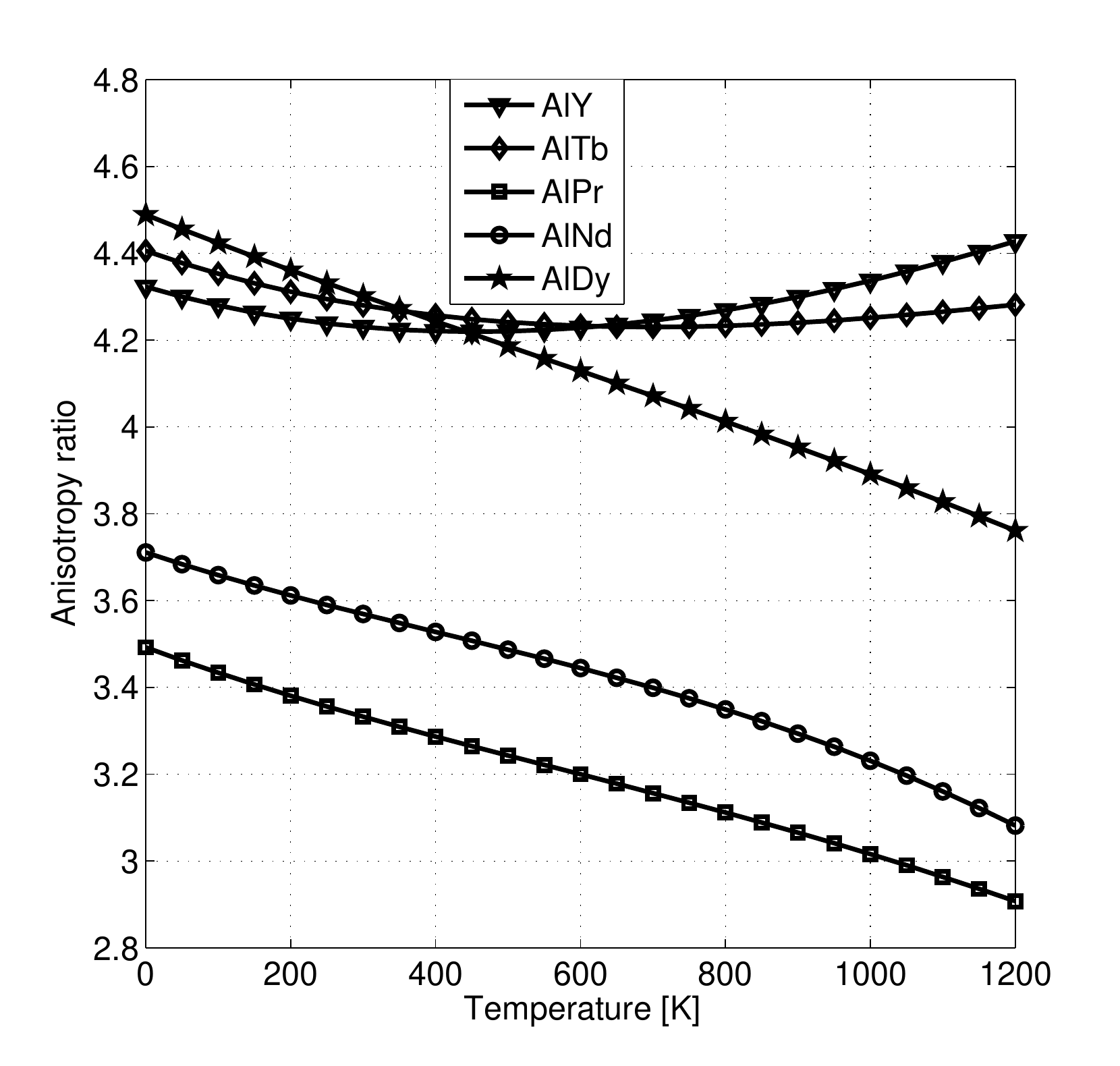}}
\caption{The anisotropy ratio of elastic constants of AlRE (RE=Y, Tb, Pr, Nd, Dy) as a function of temperature.}
\label{Aniso}
\end{figure}

The anisotropy ratio is defined as
\begin{equation}
A=\frac{C_{44}}{\mu},
\end{equation}
here we don't distinguish the isothermal and isentropic index for $C_{44}$ and $\mu$ since they have same values. At T=0K, the anisotropy ratio has the values of 4.32, 4.40, 3.49, 3.71, and 4.48 for AlY, AlTb, AlPr, AlNd, and AlDy, respectively. The elastic anisotropy are estimated
and it is shown that all the B2-AlRE compounds exhibit a substantial elastic anisotropy at 0 K.  In Figure \ref{Aniso}, we present the anisotropy ratio as a function of temperature. We can see that the anisotropy ratios of AlPr, AlNd, and AlDy decrease linearly with increasing temperature, but those of AlY and AlTb decrease initially and increase after certain temperature, i.e., there are the minimum points in the temperature curves. This phenomenon can be understood from the temperature-dependent elastic constants in Figure \ref{elastic}.  For AlPr [Figure \ref{elastic}(d)], AlNd [Figure \ref{elastic}(e)], and AlDy [Figure \ref{elastic}(f)], we find that the curves of $C_{11}^{S}$, $C_{12}^{S}$ and $C_{44}^{S}$ are nearly  parallel, i.e., almost decrease to same extent in the whole temperature range.  But for AlY [Figure \ref{elastic}(b)] and AlTb [Figure \ref{elastic}(c)], $C_{11}^{S}$ decrease to slightly larger than  $C_{12}^{S}$ and $C_{44}^{S}$, and the differences among the elastic moduli become small.

\begin{figure}
\scalebox{0.7}[0.7]{\includegraphics{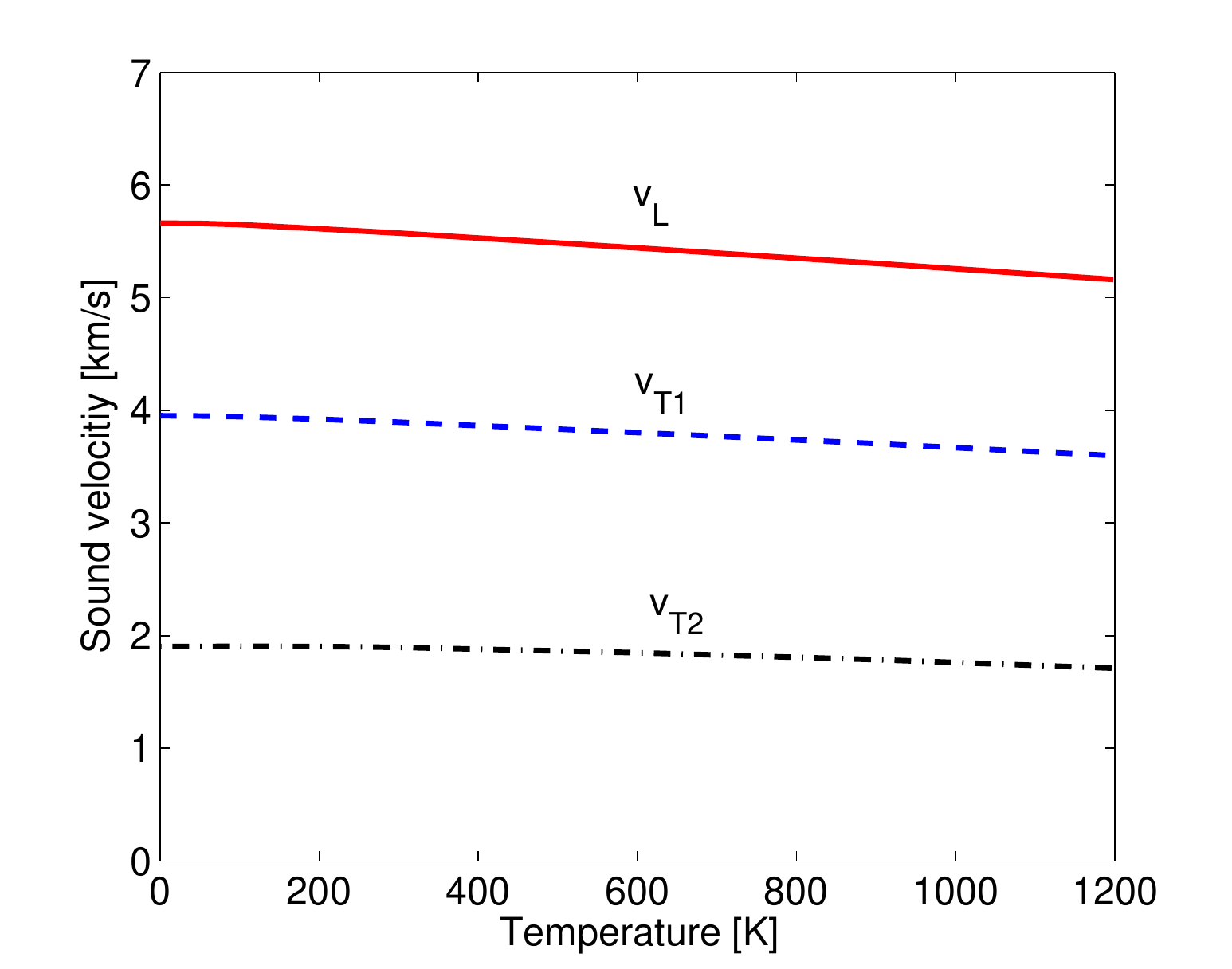}}
\caption{The temperature-dependent sound velocity along the [110] propagation direction for AlY. The solid lines denote the longitudinal mode. The dashed and dashed-dotted curves represent two transverse sound velocities polarized along [001] and [1\={1}0] directions, respectively. }
\label{soundAlY}
\end{figure}

In addition, it is convenient to discuss the sound velocities as a function of temperature  after obtaining the temperature-dependent elastic constants. From the Christoffel equation\cite{Christoffel}, sound velocities are related to the elastic constants as
\begin{equation}\label{sound}
(C^{S}_{ijkl}n_{i}n_{j}-\rho v^2 \delta_{ij})u_{i}=0,
\end{equation}
where $\mathbf{n}$ is the propagation direction, $\mathbf{u}$ is the polarization vector, $\rho$ is the density, and $v$ is the velocity. For B2 intermetallics with cubic structure, the longitudinal mode along the [110] wave propagation direction is \cite{Gulsern}
\begin{equation}\label{lsound}
\rho v_{L}^{2}=(C_{11}^S+C_{12}^S+2C_{44}^S)/2,
\end{equation}
and two transverse modes are
\begin{equation}\label{lsound}
\rho v_{T1}^{2}=C_{44}^S
\end{equation}
and
\begin{equation}\label{lsound}
\rho v_{T2}^{2}=(C_{11}^S-C_{12}^S)/2
\end{equation}
polarized along the [001] and [1\={1}0] directions, respectively. For example, we show the sound velocities of AlY propagating along the [110]  direction as a function of temperature in Figure \ref{soundAlY}. Our results show that the sound velocities decrease with increasing temperature. However, the sound velocities vary slowly in comparison with the elastic constants since the values of sound velocities are proportional to the subduplicate elastic moduli.  We also list the sound velocities at the room temperature ($T=300K$) in Table \ref{sound300}, and those data at finite temperature are very valuable for investigating the applications of AlRE intermetallics. Owing to the strong anisotropy, two transverse modes along [110] direction split obviously. In the high-symmetry direction $\Gamma-M$, the order of $v_{L}>v_{T1}>v_{T2}$ can also be obtained from the phonon dispersion curves directly.

\section{Conclusions}

In conclusions, the finite temperature elastic properties of AlRE (RE=Y, Tb, Pr, Nd, Dy) intermetallics with B2-type structures from first principles method, in which the elastic constants as a function of temperature is obtained from the combination of the static elastic constants computed from energy-strain relations by the first principles total-energy calculations and the thermal expansion obtained from the first-principles phonon theory within the quasiharmonic approach based on density-functional perturbation theory. At 0K, the equilibrium volumes, elastic constants for AlRE (RE=Y, Tb, Pr, Nd, Dy) in comparison with the previous calculated results are in excellent agreement with the previous DFT data. The overall agreement of the present results of the benchmark material Al with the ultrasonic experimental data is very good at all temperatures. The temperature dependent elastic constants of AlRE (RE=Y, Tb, Pr, Nd, Dy) intermetallics exhibit a normal behavior with temperature, i.e., decrease and approach linearity at higher temperature and zero slope around zero temperature. The anisotropy ratios of AlPr, AlNd, and AlDy decrease linearly with increasing temperature, but those of AlY and AlTb decrease initially and increase after certain temperature, i.e., there are the minimum points in the temperature curves. Finally, we have discussed the sound velocities as a function of temperature, which shows decreases with increasing temperature. However, the sound velocities vary slowly in comparison with the elastic constants since the values of sound velocities are proportional to the subduplicate elastic moduli.


 \vskip 2in

\footnotesize

\def\refname{{\large\bfseries References}}

\newpage

\begin{table}
\caption{The phonon free energies at 300K and 1000K with respect to supercell size. The unit of the energy is kJ/mol.}

\begin{tabular}{cccccc}
  \hline
  Supercell size & AlY & AlTb& AlPr& AlNd& AlDy\\
  \hline
   &  300K&  &  \\
  $2\times2\times2$ &-5.032 &-7.393 &-7.684&-7.605&-7.584\\
  $3\times3\times3$ &-4.333 &-6.566 &-6.796&-6.694&-6.576\\
  $4\times4\times4$ &-4.384 &-6.546 &-6.787&-6.682&-6.556\\
    & 1000K &  &  \\
  $2\times2\times2$ &-78.266 &-85.919&-86.789&-86.559&-85.598\\
  $3\times3\times3$ &-75.956 &-83.228&-84.348&-84.243&-83.274\\
  $4\times4\times4$ &-76.137 &-83.109&-84.124&-84.109&-83.115\\
  \hline
\end{tabular}
\label{testsize}
\end{table}

\begin{table}
\caption{{The calculated equilibrium volume $\Omega$, elastic constants $C_{ij}$, and bulk modulus $B$  for AlRE (RE=Y, Dy, Pr, Sc, Tb) intermetallics at $T=0K$ compared to previous first principles results.}}
\begin{tabular}{ccccccc}
  \hline
   & AlY & AlTb & AlPr & AlNd & AlDy  \\
  \hline
  $\Omega$ (${\AA}^3$)&46.889$^{a}$, 46.850$^{b}$&47.202$^{a}$, 47.202$^{a}$&53.157$^{a}$, 53.115$^{b}$&51.853$^{a}$, 51.812$^{b}$&46.539$^{a}$, 46.501$^{b}$\\
  $C_{11}$ (GPa)&81.26$^{a}$, 77.93$^{b}$&80.98$^{a}$, 77.13$^{b}$&72.61$^{a}$, 66.83$^{b}$&74.25$^{a}$, 69.00$^{b}$&81.40$^{a}$, 78.28$^{b}$\\
  $C_{12}$ (GPa)&54.58$^{a}$, 56.58$^{b}$&55.17$^{a}$, 56.92$^{b}$&46.82$^{a}$, 49.26$^{b}$&48.71$^{a}$, 51.09$^{b}$&56.11$^{a}$, 57.73$^{b}$\\
  $C_{44}$ (GPa)&62.79$^{a}$, 61.91$^{b}$&62.30$^{a}$, 61.17$^{b}$&47.30$^{a}$, 46.67$^{b}$&50.24$^{a}$, 49.83$^{b}$&64.00$^{a}$, 63.06$^{b}$\\
  B(GPa)        &63.47$^{a}$, 63.70$^{b}$&63.73$^{a}$, 63.66$^{b}$&55.42$^{a}$, 55.11$^{b}$&57.22$^{a}$, 57.06$^{b}$&64.54$^{a}$, 64.58$^{b}$\\
  \hline
\end{tabular}
\begin{tabular}{c}
  \leftline {${}^{a}$This work}\\
  \leftline {${}^{b}$Reference\cite{Tao1}}
\end{tabular}
\label{table}
\end{table}

\begin{table}
\caption{The sound velocities propagating along the [110] direction including one longitudinal mode and two transverse modes at T=300K are listed. The unit is $\mathrm{km/s}$. }
\begin{tabular}{ccccccc}
  \hline
   & AlY & AlTb & AlPr & AlNd & AlDy  \\
  \hline
 $v_{L}$ &5.573&4.430&4.487&4.488& 4.397\\
 $v_{T1}$&3.894&3.077&3.027&3.046& 3.067\\
 $v_{T2}$&1.894&1.488&1.658&1.612& 1.479\\
 \hline
\end{tabular}
\label{sound300}
\end{table}

\end{document}